\documentclass[pra,aps,twocolumn]{revtex4}

\usepackage{graphicx}
\usepackage{dcolumn}
\usepackage{bm}

\newcommand{\eq}[1]{Eq.~(\ref{#1})}

\begin{document}
\title{How the effective boson-boson interaction works in Bose-Fermi mixtures in
periodic geometries}
\author{G. Mazzarella}
\affiliation{Dipartimento di Fisica "G.Galilei", Universit\`{a}
degli Studi di Padova, Via F.Marzolo, 8, I-35131 Padova, Italy}
\begin{abstract}
We study mixtures of spinless bosons and not spin-polarized
fermions loaded in two dimensional optical lattices. We approach the problem of
the ground state stability within the framework of the linear response theory; by the mean of an
iterative procedure, we are able to obtain a relation for the dependence of boson-boson effective interaction on the absolute temperature of the sample.
Proceeding from such a formula,  we write down analyitical expressions for
Supersolid (SS) and Phase Separation (PS) transition temperatures, and plot the phase diagrams.
\end{abstract}
\maketitle
\section{Introduction}
The possibility to achieve very low temperatures and the
feasibility in the laboratory of implementing optical lattices
represent the proper arena where testing the validity of certain
condensed-matter theories and observing  the manifestations of
effects related to quantum mechanical statistics. Within the
interesting sinergy between quantum atom optics and condensed
matter physics, recent trends are study of the superfluid to
Mott-insulator phase transition in bosonic systems, (see
\cite{Jaksch} and \cite{Greiner}), and the striving for the
realization of a BCS-type condensate in a fermionic system
(\cite{Houbiers} and \cite{Heiselberg}). When atomic systems made
up of bosons and fermions are considered, a very rich scenario
opens up (\cite{Albus}, \cite{Illuminati2}, \cite{Salasnich}). In
particular, Bose-Fermi mixtures sympathetically cooled into their
quantum degenerate states (\cite{Truscott} and \cite{Roati})
exhibit a strong tendency towards Phase Separation (\cite{Molmer}
and \cite{Amoruso}), demixing of the bosons and the fermions, and
Supersolid (\cite{Meisel}, \cite{Lengua}, \cite{vanotterlo}).
Recently, these issues were addressed in \cite{BB}, where spinless
bosons and spin-polarized fermions confined in two dimensional
($2$D) periodic geometries were taken into account. As explained
in \cite{BB}, the effective interaction between the bosons of the
mixture plays a crucial role to the end of determining the
boundaries between the various phases of the system.\\

The dynamics underlying Phase Separation is driven by a small change
($\delta n_B$) in the bosonic density, which produces a modulation ($\delta n_F$) of the
fermionic density, related to $\delta n_B$  by  $\delta n_F \sim -U_{BF}
N(E_F)\delta n_B$, with $U_{BF}$ the on-site boson-fermion interaction
amplitude, and $N(E_F)$ the density of states at
the Fermi energy $E_F$ \cite{Mermin}. As consequence of the feedback of the
fermionic distortion $\delta n_F$, a shift of the bosonic energy $-U_{BF}^2 N(E_F)(\delta
n_B)^2/2$ occurs, thereby inducing an attraction between the bosons with
strength $U_{BF}^2 N(E_F)$.
Phase Separation emerges when the induced attraction and the intrinsic repulsion
between the bosons have the same order of magnitude \cite{BBpra}.\\

One of the peculiarities of optical lattices is their periodic
arrangement, that is the same of the crystal structure of a solid.
The phenomelogical interpretation of particle localization at
fixed sites at the base of Mott-insulator mechanism is related
just to the crystalline structure . In Supersolid, such a
structure is combined with the essence of a superfluid, i.e.
stiffnes allowing for particles current to flow without
dissipation.\\
In general to supersolids are associated two kinds of order, which usually
appear in mutual competition. These are the diagonal long-range order
(DLRO) associated with the periodic density modulation in a
crystal, and the off-diagonal long-range order (ODLRO) associated
with the phase order in the condensate \cite{Penrose}. In the strongly
interacting case, as in $^{4}$He systems,
(\cite{Andreev} and \cite{Leggett}), supersolids have been proposed
to exist and analyzed numerically in various model systems
describing interacting bosons on a lattice \cite{111213} .\\

In this paper, we consider mixtures of spinless bosons and not
spin-polarized fermions, so that a non zero $s$-wave interaction
between fermions on the same site emerges. The goal of the present
work is analyzing the effects of such an interaction on phase
diagrams of the system. Approaching the problem within the linear
response theory framework \cite{Mermin,Pines,Vignale}, we show
that it is possible to calculate in explicit way an effective
boson-boson interaction also when the on-site fermion-fermion interaction is taken into account.\\

The basic idea relies on sharing tasks between the fermions and the bosons.
In particular, the fermions are tuned through a density wave
instability establishing crystalline order (DLRO), while the
bosons provide the off-diagonal long range order (ODLRO).
The interaction between bosons and fermions, and the interaction between the
fermions, produce an additional density modulation also in the
bosonic density field, hence resulting in a SS phase. To
triggering a Density Wave (DW) instability in the fermions, the
mixed boson-fermion system is confined to two dimensions and
loaded in an optical lattice providing perfect Fermi surface
nesting at half-filling \cite{Gruner}.\\
The Supersolid transition triggered by the fermions competes with
an instability towards Phase Separation in the boson system. Because of the dimensionality
and the lattice geometry of our system, the presence of Van Hove
singularities (\cite{vanhove} and \cite{Mermin}) -well studied in BCS
superconductivity \cite{Scalapino}- strongly enhances the tendency
towards Phase Separation and produces new and interesting features
in this transition: an arbitrary weak interaction between the
bosons and the fermions, and between fermions, is sufficient to
drive the Phase Separation at low temperatures. In the following,
we investigate the instabilities towards Phase Separation and
Density Wave formation. We focus on the weak coupling limit
between the bosons and the fermions and between fermions, which
excludes a demixing in a repulsive Bose-Fermi system
along the lines discussed in \cite{Viverit}.\\

The organization of paper is as follows. In the section II, we set
the notation, and derive the model Hamiltonian. In the third
section, we display the novel result consisting in the analytic
formula for the effective boson-boson interaction as a function of
the temperature, and discuss the competition between PS and SS; we
derive the Phase separation and Density-Wave transition
temperatures. In the last section, we write down our conclusions
and comment about future perspectives of the topic.

\section{The model Hamiltonian}
The microscopic Hamiltonian for interacting spinless bosons and
not spin-polarized fermionic atoms subject to an optical lattice
reads $\hat{H}=\hat{H}_B+\hat{H}_F+\hat{H}_{int}$ (here and in the
following, $\alpha=B,F$ and $\sigma={\uparrow,\downarrow}$ denote
the atomic species and the spin, respectively) \cite{Albus,
Fetter}
\begin{eqnarray}
\label{fullhamiltonian}
&&\hat{H}_{\alpha} =\int d \vec{r} \hat{\Psi}_{\alpha}^{\dagger}\bigg(
-\frac{\hbar^{2}}{2
m_{\alpha}}\nabla^2+V_{\alpha} \bigg)\hat{\Psi}_{\alpha},\nonumber\\
&& \hat{H}_{int}=\int d \vec{r} \bigg[\frac{g_{BB}}{2}\hat{\Psi}_{B}^{\dagger}\hat{\Psi}_{B}^{\dagger}
\hat{\Psi}_{B}\hat{\Psi}_{B}\nonumber\\
&+&g_{BF}(\hat{\Psi}_{B}^{\dagger}\hat{\Psi}_{B}
\hat{\Psi}^{\dagger}_{F,\uparrow}\hat{\Psi}_{F,\uparrow}+\hat{\Psi}_{B}^{\dagger}\hat{\Psi}_{B}
\hat{\Psi}^{\dagger}_{F,\downarrow}\hat{\Psi}_{F,\downarrow})\nonumber\\
&+& \frac{g_{FF}}{2}\sum_{\sigma^{'} \neq \sigma}
\hat{\Psi}^{\dagger}_{F,\sigma}
\hat{\Psi}^{\dagger}_{F,\sigma^{'}}\hat{\Psi}_{F,\sigma^{'}}\hat{\Psi}_{F,\sigma}\bigg]
.\end{eqnarray} Here, we assume a repulsive interaction between
the bosons: $g_{BB}=4\pi\hbar^{2}a_{BB}/m_{B}>0$, where $a_{BB}$
is associated the $s$-wave scattering length. The boson-fermion
interaction strength $g_{BF}=4\pi\hbar^{2}a_{BF}/m_{R}$ is assumed
to be the same for both the spin configurations; here, $a_{BF}$ is
the boson-fermion $s$-wave scattering length and
$m_{R}=m_{B}m_{F}/(m_{B}+m_{F})$ is the reduced mass. Finally,
$g_{FF}=2\pi\hbar^{2}a_{FF}/m_{F}$ with $a_{FF}$ the
fermion-fermion $s$-wave scattering length. In the following we
always assume both $a_{BF}>0$ and $a_{FF}>0$.

The optical lattice with wave length $\lambda$ provides an
$a=\lambda/2$- periodic potential for the bosons and fermions with
$\displaystyle{V_{B,F}(x,y)=V_{B,F}\big(\sin^2 \frac{\pi
x}{a}+\sin^2 \frac{\pi y}{a}\big)}$ \cite{Jessen}, where the
lattice depth $V_{F}$ is assumed the same for both the spin
configurations. In the experiments, the $2$D setup is realized by
generating an anisotropic three-dimensional optical potential
$\displaystyle{V_{B,F}(x,y)=V_{B,F}\big(\sin^2 \frac{\pi
x}{a}+\sin^2 \frac{\pi y}{a}\big)+V^{z}_{B,F}\sin^2 \frac{\pi
z}{a_z}}$ with $V^{z}_{B,F} \gg V_{B,F}$; then
the interplane hopping is quenched.\\
Due to the strong localization around each lattice site
$\vec{r}_i$, the annihilation (creation) bosonic and fermionic
field operators $\hat{\Psi}_{\alpha}$
($\hat{\Psi}^{\dagger}_{\alpha}$) may be expanded in terms of the
Wannier functions $w^{l}_{\alpha}(\vec{r}-\vec{r}_i)$, with $l$
the Bloch band index \cite{Jessen}
\begin{equation}
\label{wannierexpansionB} \hat{\Psi}_{\alpha}(\vec{r}) \, = \,
\sum_{i,l} \hat{a}^{l}_{i} w^{l}_{\alpha}(\vec{r}-\vec{r}_i) \; ,
\end{equation}
where $\hat{a}^{l}_{i}$ is the bosonic ($\hat{b}^{l}_i$) or
fermionic ($\hat{c}^{l}_{\sigma,i}$) annihilation operator acting
on a particle at the \mbox{$i$th} lattice site and in the $l$th
Bloch band. For a strong optical lattice
$V_{B,F}>E^{r}_{B,F}=2\hbar^2\pi^2/\lambda^2 m_{B,F}$ the
restriction to the lowest Bloch band ($l=0$) is justified
\cite{Jaksch}. The translationally invariant lattice version of
Hamiltonian (\ref{fullhamiltonian}) reads
\begin{widetext}
\begin{equation}
\label{latticeh1} \hat{H} \, = \, -J_{B}\sum_{<i,j>}
\,\hat{b}^{\dagger}_{i} \hat{b}_{j} \,
-J_F\sum_{<i,j>,\sigma} \,\hat{c}^{\dagger}_{i,\sigma}
\hat{c}_{j,\sigma}+ \, \frac{U_{BB}}{2}\sum_{i}
\hat{n}_{i}(\hat{n}_{i}-1)+ U_{BF}\sum_{i,\sigma}
(\hat{n}_{i}\hat{m}_{i,\sigma})+ U_{FF}\sum_{i}
(\hat{m}_{i,\uparrow}\hat{m}_{i,\downarrow})+\delta\sum_{i}(\hat{m}_{i,\uparrow}-\hat{m}_{i,\downarrow})
\; ,\end{equation}
\end{widetext}
where we have omitted the band index $l=0$. The symbol $<i,j>$
denotes couples of nearest-neighbor lattice sites, and $\delta$
the imbalance between spin-up and spin-down fermions;
$\hat{n}_{i}=\hat{b}^{\dagger}_{i}\hat{b}_{i}$ and
$\hat{m}_{i,\sigma}=\hat{c}^{\dagger}_{i,\sigma}\hat{c}_{i,\sigma}$
are the number operators for bosons and fermions with spin
$\sigma$ at the $i$th site, respectively. The boson-boson,
boson-fermion, and fermion-fermion interaction amplitudes are
$U_{BB}=g_{BB}\int d \vec{r}|w_{B}(\vec{r})|^4 $,
$U_{BF}=g_{BF}\int d \vec{r}|w_{B}(\vec{r})|^2
|w_{F}(\vec{r})|^2$, and $U_{FF}=g_{FF}\int d \vec{r}
|w_{F,\uparrow}(\vec{r})|^2 |w_{F,\downarrow}(\vec{r})|^2$,
respectively \cite{Jaksch,Albus,BB}.

Among the adavantages provided by optical lattices, there is the
possibility of tuning  the Hamiltonian parameters in such a way to
realize different interaction regimes.  Here we focus on weak
coupling regime, which takes place when $\lambda_{BF}=U_{BF}^2
N_0/U_BB<<1$ ($N_0=1/2\pi^2 J_F$) and $t_B=8 J_B/n_{B} U_{BB} >>
\lambda_{BF} $, where $n_B$ is the
bosonic filling factor. \\
The parameters involved in the Hamiltonian (\ref{latticeh1})
are related to characteritistic quantities of the optical potential according to
\begin{widetext}
\begin{eqnarray}
\label{amplitudes}
J_{B,F}=(4\sqrt{\pi})E^{r}_{B,F}V_{B,F}^{3/4}\exp(-2\sqrt{V_{B,F}});\nonumber\\
\frac{U_{BF}}{E_{F}^{r}}=8\sqrt{\pi}\frac{1+m_F/m_B}{1+\sqrt{V_F/V_B}}\frac{a_{BF}}{\lambda\gamma}(V_{F}^z)^{1/4}(V_{F})^{1/2};\nonumber\\
\frac{U_{BB}}{E_{B}^{r}}=4\sqrt{2\pi}\frac{a_{BB}}{\lambda\gamma}(V_{B}^z)^{1/4}(V_{B})^{1/2};\nonumber\\
\frac{U_{FF}}{E_{F}^{r}}=4\sqrt{2\pi}\frac{a_{FF}}{\lambda\gamma}(V_{F}^z)^{1/4}(V_{F})^{1/2}.\end{eqnarray}
\end{widetext}
where $\gamma=2 a_z/\lambda$. The hopping and the atom-atom
interaction amplitudes (\ref{amplitudes}) are evaluated by
extending to our case the calculations performed in
\cite{Jaksch,Albus,BB}.\\

By exploiting the discrete Fourier transform  of $\hat{a}_{i}$ and of its
Hermitian conjugate, the Hamiltonian (\ref{latticeh1}) may be written in the
momenta space
\begin{widetext}
\begin{eqnarray}
\label{lattice h}
\hat{H}=\sum_{\vec{k}}\big[\epsilon_{B,\vec{k}}\hat{b}^{\dagger}_{\vec{k}}\hat{b}_{\vec{k}}+(\epsilon_{\uparrow,\vec{k}}-\delta)
\hat{c}_{\uparrow,\vec{k}}^{\dagger}\hat{c}_{\uparrow,\vec{k}}+(\epsilon_{\downarrow,\vec{k}}+\delta)\hat{c}_{\downarrow,\vec{k}}^{\dagger}\hat{c}_{\downarrow,\vec{k}}\big]+\nonumber\\
\frac{1}{M^2}\sum_{\vec{k}}\big[\frac{U_{BB}}{2}\hat{n}_{B,\vec{k}}\hat{n}_{B,-\vec{k}}+U_{BF}\hat{n}_{B,\vec{k}}\hat{m}_{F,-\vec{k}}+
U_{FF}\hat{m}_{\uparrow,\vec{k}}\hat{m}_{\downarrow,-\vec{k}}]
,\end{eqnarray}
\end{widetext} where $M$ is the number of lattice sites in $x$ ($y$) direction. We assume the wave vector $\vec{k}$ be
restricted to the first Brillouin zone: $k_x\in [-\pi/a,\pi/a]$,
$k_y\in [-\pi/a,\pi/a]$. The density number operators are
$\hat{n}_{B,\vec{k}}=\sum_{\vec{p}}\hat{b}_{\vec{k}+\vec{p}}^{\dagger}\hat{b}_{\vec{p}}$,
$\hat{m}_{\sigma,\vec{k}}=\sum_{\vec{p}}\hat{c}_{\sigma,\vec{k}+\vec{p}}^{\dagger}\hat{c}_{\sigma,\vec{p}}$
\cite{Vignale};
the bosonic and fermionic dispersion relations $\epsilon_{B,\vec{k}}$ and
$\epsilon_{\sigma,\vec{k}}$ read
$\epsilon_{B,\vec{k}}=- 2J_{B}(2-\cos(k_x a)-\cos(k_ya))$ and
$\epsilon_{\sigma,\vec{k}}=-2J_{F,\sigma}(\cos(k_x
a)-\cos(k_ya))$ \cite{Fetter}. \\

\section{Phase separation versus Supersolid}
In this section, we want to gain analytical insight into stability of  the mixture ground state.
As explained in \cite{BB}, to achieve such a goal, the starting point is the
derivation of an effective Hamiltonian for the bosons alone, by keeping in mind
that fermions enters the description of our system via a modified interaction
between the bosons themselves. Within linear response theory \cite{Mermin,Vignale} the boson
density $n_{B}(\vec{q})$ (for the sake of simplicity, we
refer to $n_{B} (\vec{q}) (m_{\sigma} (\vec{q}))$ as to the
induced perturbation) drives the fermionic system trough the
following modulation of the density
\begin{eqnarray}
\label{lrt}
<m_{\sigma,\vec{q}}>=\chi_{\sigma}(U_{BF}n_{B,\vec{q}}+U_{FF}<m_{-\sigma,\vec{q}}>).\nonumber\\
\end{eqnarray}
The function $\chi_{\sigma}$, for the $\sigma$ component of the spin, is
the Lindhard function depending upon the absolute temperature $T$
and on the wave vector $\vec{q}$ \cite{Mermin,Vignale}:
\begin{eqnarray}
\label{Lindhard} \chi_{\sigma}(T,\vec{q})=\frac{1}{2}\int
\frac{d\vec{k}}{v_0}\frac{f(\epsilon_{\sigma,\vec{k}})-f(\epsilon_{\sigma,\vec{k}+\vec{q}})}{\epsilon_{\sigma,\vec{k}}-\epsilon_{\sigma,\vec{k}+\vec{q}}},
\end{eqnarray}
where $v_0=(2\pi/a)^{2}$ is the volume of the first Brillouin
zone; the integration is performed over this region. The
temperature $T$ enters via the Fermi distribution function
$\displaystyle{f(\epsilon_{\sigma,\vec{k}})=1/\big((1+\exp(\frac{\epsilon_{\sigma,\vec{k}}-\mu_F}{T})\big)}$,
with $\mu_F$ the chemical potential of the fermionic atoms.\\
We analyze the behavior of the system for temperatures well below
the the superfluid transition temperature $T_{KT}$
(Kosterlitz-Thouless). Hence, the mixture is enhanced in a
sufficiently low temperature regime so that
the fermionic chemical potential $\mu_F$ my be safely identified with the Fermi
energy $E_F$.\\

Proceeding form the Hamiltonian (\ref{lattice h}), we integrate
out the fermionic freedom degrees following the same path as in
\cite{BB}. Within the procedure of tracing out the fermions, we
treat the spin-up component independently from the other, i.e. we
perform a mean-field approximation. We get the Hamiltonian
\begin{widetext}
\begin{eqnarray}
\label{effectivehamiltonian}
\hat{H}^{int}_{eff}=\frac{1}{M^2}\big[\sum_{\vec{k}}\frac{U_{BB}}{2}\hat{n}_{B,\vec{k}}\hat{n}_{B,-\vec{k}}+
U_{BF}(\hat{n}_{B,\vec{k}}<\hat{m}_{\uparrow,-\vec{k}}>+\hat{n}_{B,\vec{k}}<\hat{m}_{\downarrow,-\vec{k}}>)+
U_{FF}<\hat{m}_{\uparrow,\vec{k}}><\hat{m}_{\downarrow,-\vec{k}}>\big],\end{eqnarray}\end{widetext}
which describes the effective interaction between the bosons of the mixture.
By employing the rules summarized in Eq. (\ref{lrt}) in the Hamiltonian
(\ref{effectivehamiltonian}), we obtain
\begin{widetext}
\begin{eqnarray}
\label{recursion}
\hat{H}^{int}_{eff}=\frac{1}{M^2}\sum_{\vec{k}}\big[U_{BF}\big(\hat{n}_{B,\vec{k}}(\chi_{\uparrow}U_{BF}\hat{n}_{B,-\vec{k}}+\chi_{\uparrow}U_{FF}<\hat{m}_{\downarrow,-\vec{k}}>\big)+\nonumber\\
U_{BF}\big(\hat{n}_{B,\vec{k}}(\chi_{\downarrow}U_{BF}\hat{n}_{B,-\vec{k}}+\chi_{\downarrow}U_{FF}<\hat{m}_{\uparrow,-\vec{k}}>\big)+\nonumber\\
U_{FF}\big(\chi_{\uparrow}U_{BF}\hat{n}_{B,\vec{k}}+\chi_{\uparrow}U_{FF}<\hat{m}_{\downarrow,\vec{k}}>)
(\chi_{\downarrow}U_{BF}\hat{n}_{B,-\vec{k}}+\chi_{\downarrow}U_{FF}<\hat{m}_{\uparrow,-\vec{k}}>)\big]
.\end{eqnarray}
\end{widetext}
We exploit the rules (\ref{lrt}) in Eq. (\ref{recursion}) in
iterative way. Then, the Hamiltonian (\ref{recursion}) can be
expressed as an expansion in series of powers of
$|\chi_{\sigma}U_{FF}|$; we have verified that this last quantity
is much smaller than one. We assume to be in a situation in which
the spin-up population is equal to the spin-down one ($\delta
=0$), and in which the fermions are in half-filling configuration,
$E_F=0$ and $m_{\uparrow}=m_{\downarrow}=1/4$. In such a situation
the series associated to the Hamiltonian (\ref{recursion}) may be
summed. We employ the fact that when $\delta =0$, is
$\epsilon_{\uparrow}=\epsilon_{\downarrow} \equiv \epsilon_F$ and
then $\chi_{\uparrow}=\chi_{\downarrow} \equiv \chi$, with
\begin{eqnarray}
\label{Lindhardbalance} \chi(T,\vec{q})=\frac{1}{2}\int
\frac{d\vec{k}}{v_0}\frac{f(\epsilon_{F,\vec{k}})-f(\epsilon_{F,\vec{k}+\vec{q}})}{\epsilon_{F,\vec{k}}-\epsilon_{F,\vec{k}+\vec{q}}},
\end{eqnarray}
Together with these two last properties, we use the symmetry
$\epsilon_{F,\vec{k}}=\epsilon_{F,-\vec{k}}$; then, the effective
boson-boson interaction (\ref{effectivehamiltonian}) reads
\begin{widetext}
\begin{eqnarray}
\label{recursionbis}
\hat{H}^{int}_{eff}=\frac{1}{2M^2}\sum_{\vec{k}}\bigg[U_{BB}+2\chi(T,\vec{q})U_{BF}^{2}\big(\frac{1}{1-\chi(T,\vec{q})
U_{FF}}+\frac{U_{FF}}{(\chi(T,\vec{q})^2
U_{FF}-1)^2}\big)\bigg]\hat{n}_{B,\vec{k}}\hat{n}_{B,-\vec{k}}
.\end{eqnarray}\end{widetext} The effective boson-boson
interaction $U_{eff}(T)$ depends on the temperature according the
formula
\begin{eqnarray}
\label{Ueff}
&&U_{eff}(T)=U_{BB}+2\chi(T,\vec{q})U_{BF}^{2}\big(\frac{1}{1-\chi(T,\vec{q})
U_{FF}}\nonumber\\
&+&\frac{\chi(T,\vec{q})U_{FF}}{(\chi(T,\vec{q})
U_{FF}-1)^2}\big).\end{eqnarray} The novelty of the present work
relies on Eq. (\ref{Ueff}). Such a formula represents a very
useful tool for investigating in analytical way the instability
related to the Van Hove singularity and the one associated to the
Density Wave. In particular, we are interested in calculating the
temperatures of transition to PS and to DW phases.\\
To this end, let us focus on fermionic dispersion relation
$\epsilon_{F,\vec{k}}=-2 J_{F}[\cos(k_xa)+\cos(k_ya)]$. By
analyzing $\epsilon_{F,\vec{k}}$, we realize that the Lindhard
function (\ref{Lindhardbalance}) exhibits two logarithmic
singularities. These two singularities give rise to instabilities
in the system. In particular, the singularity at $\vec{q}=\vec{0}$
induces an instability towards Phase Separation. On the other
hand, in correspondence of the wave vector
$\vec{k}_{DW}=(\pi/a,\pi/a)$ -which joints the saddle points (SP)
$\vec{k}_{SP}=(0,\pm \pi/a),(\pm \pi/a,0)$ of the fermionic
dispersion relation- the fermionic energy vanishes. The wave
vector $\vec{k}_{DW}$ drives a
Density Wave, responsible for the Supersolid order. \\
Let us focus on the logarithmic Van Hove singularity, and analyze
the density states. Due to the balance between the spin-up and the
spin-down populations, we have that
$N_{\uparrow}(\epsilon)=N_{\downarrow}(\epsilon)\equiv
N(\epsilon)$. Within the tight-binding regime, $N(\epsilon)$ reads
\cite{BB,Mermin}
\begin{eqnarray}
\label{densityofstates}
N(\epsilon)=\frac{1}{2}N_{0}K[\sqrt{1-\frac{(\epsilon \pm
\delta)^2}{16 J_F^2}}]\sim \frac{N_0}{2} \ln|\frac{16
J_F}{(\epsilon \pm \delta)}|,\end{eqnarray} with $K[k]$ the
complete elliptic integral of the first kind \cite{Abr}.\\
We write the response function (\ref{Lindhardbalance}) in the
energy space \cite{Mermin}, and analyze it at $\vec{q}=\vec{0}$
and in the limit of zero temperature
\begin{eqnarray}
\label{Lindhardenergy} \chi(T \to 0,0)=\int_{0}^{8J_F} d\epsilon
N(\epsilon)_{\delta=0}\partial_{\epsilon}
f(\epsilon)_{E_F=0}\nonumber\\
\sim -\frac{N_0}{2}\ln(\frac{16 c_1 J_F }{T}),\end{eqnarray} where
$8 J_F$ is the bandwidth, and the subscripts $\delta=0$ and
$E_F=0$ denote the absence of population imbalances and the
half-filled band situation, respectively; $c_1=(2\exp(C))/\pi
\approx 1.13$ with $C$ the Eulero constant ($\approx 0.577...$).
The coupling between the bosons and the fermions induces an
attraction between the bosons, and the effective long distance
scattering parameter takes the form
\begin{eqnarray}
\label{scatteringlongdistance}
U_{eff}(T)=U_{BB}+2\chi(T,0)U_{BF}^{2}\big(\frac{1}{1-\chi(T,0)
U_{FF}}+\nonumber\\\frac{\chi(T,0)U_{FF}}{(\chi(T,0)
U_{FF}-1)^2}\big).
\end{eqnarray}
The thermodynamic stability of a superfluid condensate at low
temperatures requires a positive effective interaction, i.e.
$U_{eff}(T)>0$ as commented in \cite{Abrikosov}. The condition
$U_{eff}(T_{PS})=0$ defines the critical temperature $T_{PS}$ for
Phase Separation
\begin{widetext}
\begin{eqnarray}
\label{PStemperature} T_{PS}=16 c_1 J_{F} \exp[\bigg(\frac{2}{N_0
U_{FF}-(1+\sqrt{1-\frac{2 N_0
U_{FF}}{\lambda_{BF}}})\lambda_{BF}}\bigg)].\end{eqnarray}\end{widetext}
\begin{widetext}
Let us focus, now, on the instability in the system triggered by
$\vec{k}_{DW}$. By exploiting the symmetry
$\epsilon_{F,\vec{q}+\vec{k}_{DW}}=-\epsilon_{F,\vec{q}}$, the
energy representation of the response function, in the energy
representation, reads
\begin{eqnarray}
\label{Lindharddensitywave} \chi(T,\vec{k}_{DW})=\int_{0}^{8 J_F}
d \epsilon
N(\epsilon)\frac{\tanh(\epsilon/2 T)}{-2\epsilon}\nonumber\\
\sim-\frac{N_0}{4} \big(\ln(\frac{16 c_1
J_F}{T})\big)^{2}.\end{eqnarray} Within Bogoliubov theory, the
bosonic quasiparticle spectrum is \cite{Abrikosov}
\begin{eqnarray}
\label{spectrum}
E_{B}(\vec{q})=\sqrt{\epsilon_B(\vec{q})^2+2n_B\epsilon_B(\vec{q})[U_{eff}(T)]}.\end{eqnarray}
The induced attraction between the bosons reduces the energy of
quasiparticle providing a roton minimum at $\vec{k}_{DW}$, which
vanishes at critical temperature $T_{DW}$ given by
\begin{eqnarray}
\label{DWtemperature} T_{DW}=16 c_1 J_F \exp\bigg[-2
\sqrt{\frac{1}{-N_0
U_{FF}+\frac{2\lambda_{BF}}{2+t_B}+\frac{2\lambda_{BF}\sqrt{\frac{-N_0
(2+t_B)U_{FF}+\lambda_{BF}}{\lambda_{BF}}}}{2+t_B}}}\bigg].\end{eqnarray}\end{widetext}
By performing the limit $U_{FF} \to 0$ in Eq.
(\ref{PStemperature}) and in Eq. (\ref{DWtemperature}) we are able
to reproduce the temperatures signing the Phase Separation and
Density Wave calculated in \cite{BB}.\\

The fermion-fermion interaction plays a crucial role in
determining the order characterizing the ground state of the
mixture. The interaction between the fermions influences the
behavior of the system via the expression of the effective
boson-boson interaction, as summarized in \eq{Ueff}. A non
vanishing repulsive fermion-fermion interaction acts in such a way
to rising a barrier of potential, which prevents system to achieve
Phase Separation. This reflects in a mechanism which advantages
the Supersolid order, as shown in the plots reported in Fig. 1
($U_{FF}>0$) and in Fig. 2 ($U_{FF}=0$), that are represented for
a mixture made up of bosonic atoms of $^{87}$Rb and of fermionic
atoms of $^{40}$Rb in a square optical lattice with $\lambda$
equal to $830$nm. From these figures, we see that the PS and SS
lines cross in a certain point, characterized by a critical depth.
We observe that the PS region of Fig. 1, plotted for $U_{FF}>0$,
is smaller than the same region of Fig. 2 obtained with
$U_{FF}=0$. The difference between the magnitudes of the two PS zones may be interpreted as
a signature of the role of the finite fermion-fermion interaction.
The amount of the shift of a PS region with respect to the other might be used as an indirect measure of the fermion-fermion $s$-wave scattering
length.\\
The role of the temperature is important in determining the kind
of order that in the system establishes as well. In fact, when the
optical lattice depth $V_{F}$ is greater than its critical value,
the greater is the temperature the more the Phase Separation is
dominant in the system; below the critical depth, the behavior is
reversed. The comparison between the absence of the on-site
fermion-fermion interaction and the case in which such an
interaction is present, is shown in Fig. 3. Here, we observe that
when the fermions of mixture interact with each other, the SS
region is wider with respect to $U_{FF}=0$ case. Then, we may
deduce that the Fermi-Fermi interaction acts in such a way to make
more important the Supersolid order.


\begin{figure} \centering
\includegraphics[width=8cm]{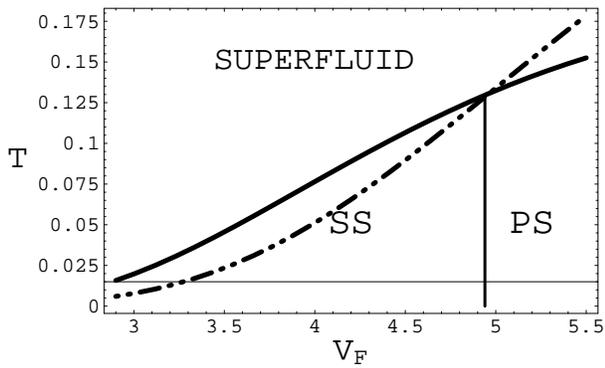}
\caption{On the horizontal axis the fermionic optical depth in
fermionic recoil energy units; on the vertical axis the
temperature normalized to the fermionic recoil energy. The
dot-dashed and continuous  lines are the PS and DW temperatures,
respectively, when $U_{FF} > 0$. The curves are plotted for
$a_{FF}=104.8$ a$_0$, with a$_0$ the Bohr radius, $\gamma=3$,
$n_B=3/2$, $V_{F}^{z}=20$,$V_{B}/V_{F}=V_{B}^{z}/V_{F}^{z}=7/3$.}
\label{globalsecond}
\end{figure}
\begin{figure}\centering
\includegraphics[width=8cm]{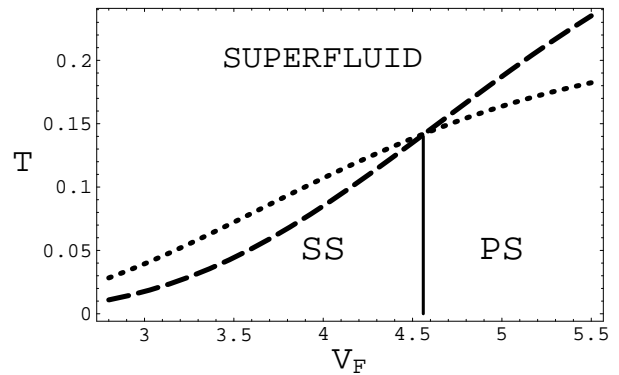}
\caption{The dashed and dotted lines are the PS and DW
temperatures, respectively, when $U_{FF}=0$. Here is, clearly,
$a_{FF}=0$, and the other quantities assume the same values as in
Fig. 1.} \label{globalfirst}
\end{figure}
\begin{figure}\centering
\includegraphics[width=8cm]{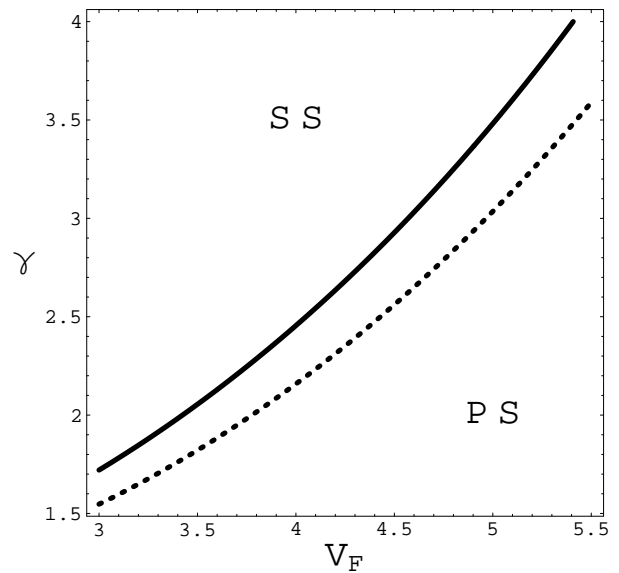}
\caption{The black continuous line represents the separation
between SS (the points above the line) and the PS (the points
below the line) when $a_{FF}=0$; the dotted line is plotted in
with $a_{FF}=104.8$ a$_0$. The other quantities assume the same
values as in Fig. 1.} \label{Pd}
\end{figure}

\section{Conclusions}
In this paper, we have investigated the ground state of a
Bose-Fermi mixture loaded in a two-dimensional periodic geometry,
and made up of spinless bosons and not spin-polarized fermions. We
have analyzed the competition between Phase Separation and
Supersolid orders. This analysis was carried out within the
framework of the linear response theory. By employing an iterative
technique, we have obtained an analytical expression for the
effective boson-boson interaction as a function of the absolute
temperature of the sample. We have performed our study in absence
of fermionic population imbalances and in presence of
boson-fermion and fermion-fermion repulsive interactions. We have
studied the phase diagram of the system by stressing the role of
the fermion-fermion interaction in determining the kind of order
sustained by the Bose-Fermi mixture Hamiltonian. We have explained
the changes that such an interaction introduces in the behavior of
the system with respect to the case in which the fermions do not
interact on the same lattice site. We have discussed the nice
interplay between the critical value of the optical lattice depth
for the fermions and the temperature of the sample in
establishing the kind of order in the ground state of the mixture.\\

The study of such a topic opens up a bunch of very exciting
possibilities. In particular, a very interesting relationship
could be established between these issues and the ones related to
propagation of the zero sound waves both in homogeneous
\cite{Pines} and in trapped \cite{Padova} systems of ultracold
atoms; for these kind of problems, in fact, the linear response
theory is exploited as well.

In a forthcoming paper, we wish to accomplish the task of
analyzing the case of imbalance between the two fermionic
populations and different sign of boson-fermion and
fermion-fermion interactions, by employing a suitable modification
of the iterative method displayed in the present work.\\

This work has been partially supported by Fondazione CARIPARO. Discussions with S.M. Giampaolo are acknowledged.

\end{document}